\documentclass[lettersize,journal]{IEEEtran}
\usepackage{amsmath,amssymb,amsfonts}
\usepackage[linesnumbered,ruled,vlined]{algorithm2e}
\usepackage[caption=false,font=normalsize,labelfont=sf,textfont=sf]{subfig}
\usepackage{textcomp}
\usepackage{stfloats}
\usepackage{url}
\usepackage{verbatim}
\usepackage{graphicx}
\usepackage{hyperref}
\usepackage{cite}
\usepackage{fancyhdr}
\usepackage{tabularray}
\usepackage[inline]{enumitem}
\usepackage{xcolor}
\usepackage{comment}
\usepackage{diagbox}
\usepackage{tikz}
\begin{document}

\title{Beacon: LLM Multi-Agent Driven Hardware Design Space Exploration for Heterogeneous Multi-Chiplet Deep Learning Accelerators}

\author{Boyu~Li,
Zongwei~Zhu\textsuperscript{$\ast$},
    Qianyue~Cao,
    Xi~Li, and
 Xuehai~Zhou,~\IEEEmembership{Member,~IEEE}
 
\thanks{This work was supported by the National Natural Science Foundation of China (62572455). \emph{(Corresponding author: Zongwei Zhu.)}}

\thanks{Boyu Li, Qianyue Cao, and Xuehai Zhou are with the School of Computer Science and Technology, University of Science and Technology of China, Hefei 230026, China and Suzhou Institute for Advanced Research, University of Science and Technology of China, Suzhou 215123, China  (e-mail: llbbyy@mail.ustc.edu.cn; cqy\_1999@mail.ustc.edu.cn; xhzhou@ustc.edu.cn).}

\thanks{Zongwei Zhu and Xi Li are with the School of Software Engineering, University of Science and Technology of China, Hefei 230026, China and Suzhou Institute for Advanced Research, University of Science and Technology of China, Suzhou 215123, China (e-mail: zzw1988@ustc.edu.cn; llxx@ustc.edu.cn).}

}

\maketitle

\begin{abstract}
Heterogeneous multi-chiplet accelerators allow chiplets to be configured independently to better match different operator characteristics and improve inference efficiency. However, heterogeneity makes simulator evaluation expensive, limiting the number of iterations affordable for hardware design space exploration (HW-DSE). Mainstream data-driven methods rely mainly on final metrics and a few predefined states, and require many search iterations to implicitly learn the relationships between input parameters and optimization objectives, making them less effective in this setting. In practice, evaluators also generate detailed reports on execution timelines, resource utilization, memory accesses, and communication behavior. Large language models (LLMs) can combine domain knowledge with these reports to explicitly identify bottleneck locations, degradation causes, and parameter adjustment directions, thereby improving each design decision under limited iteration budgets. Based on this observation, we propose Beacon, a report-driven LLM multi-agent framework for heterogeneous multi-chiplet HW-DSE. Beacon employs hierarchical agents for bottleneck localization, root-cause diagnosis, and hardware candidate generation, together with an Analysis Toolbox and RAG memory for closed-loop search. Under the same limited iteration budget, Beacon reduces the composite latency-energy-monetary-cost objective by 25.1\%--93.5\% compared with random search, Bayesian optimization, and reinforcement learning.
\end{abstract}


\section{Introduction}
\label{sec:introduction}

As deep learning models continue to grow in scale and complexity, particularly in the era of large language models (LLMs), conventional technology scaling alone is increasingly insufficient to meet the computational demands of model inference. Monolithic accelerators are constrained by wafer yield and reticle limits, making further scaling through larger dies increasingly difficult \cite{single-chip-limit}. Multi-chiplet integration provides a scalable path for large-model inference. Heterogeneous multi-chiplet accelerators further allow individual chiplets to adopt different dataflows, compute resources, and memory configurations, enabling finer-grained adaptation to different inference stages and operator characteristics and thereby improving inference efficiency \cite{moham,scar}.

However, heterogeneity also significantly increases the cost of hardware design space exploration (HW-DSE) \cite{hetero-chiplet}. Each HW-DSE iteration typically consists of two steps: design decision and simulator evaluation. The design decision itself takes relatively little time, whereas simulator evaluation requires joint workload mapping search, chiplet-level simulation, and inter-chiplet communication modeling, and therefore dominates the runtime of each iteration \cite{varipar}. This issue is more pronounced for heterogeneous multi-chiplet designs because workload performance must be evaluated on different chiplet configurations, causing the evaluation time to increase linearly with the number of heterogeneous chiplets. Even on a high-performance multicore server, a complete evaluation typically takes tens of minutes, limiting one HW-DSE run to only tens of iterations \cite{compass}. Consequently, mainstream HW-DSE methods often perform poorly in this setting.

Current HW-DSE decision methods primarily adopt data-driven approaches, including offline surrogate modeling and online methods based on reinforcement learning and Bayesian optimization (BO) \cite{mind-mappings,chiplet-gym,confuciux,compass}. These methods often require thousands of iterations to converge to a satisfactory design point, making them unsuitable for heterogeneous multi-chiplet HW-DSE with a limited iteration budget. We observe that this high iteration demand largely stems from the limited information available to each design decision. Data-driven methods typically treat the evaluator as a black box and make decisions based mainly on the final metrics produced by each evaluation and a small set of predefined state variables. Consequently, they must rely on repeated trials to implicitly capture the relationships between input parameters and final objectives.

LLMs provide a new opportunity to extract more decision-relevant information from each evaluation~\cite{chateda,layoutcopilot,atelier}. In addition to final metrics, modern accelerator evaluators produce fine-grained reports containing execution timelines, per-layer latency and energy, resource utilization, mapping behavior, DRAM accesses, and inter-chiplet communication. Traditional decision methods usually represent each evaluation as a search sample consisting of the hardware configuration, predefined states, and final metrics, without explicitly modeling the diagnostic information in the reports, such as bottleneck locations, causes of performance degradation, and interactions among hardware parameters. Consequently, when a candidate configuration performs poorly, traditional methods cannot readily determine from a single evaluation whether an individual parameter is ineffective or whether the degradation is caused by an inappropriate combination of parameters. Instead, they must gradually learn the statistical relationships between parameters and objectives from subsequent samples. In contrast, an LLM can combine hardware domain knowledge with evaluator reports and transform each evaluation into a diagnostic result that identifies the bottleneck, attributes its root cause, and suggests parameter adjustment directions, thereby providing more direct guidance for the next search decision.

\begin{figure}[ht]
\centering
\includegraphics[width=0.48\textwidth]{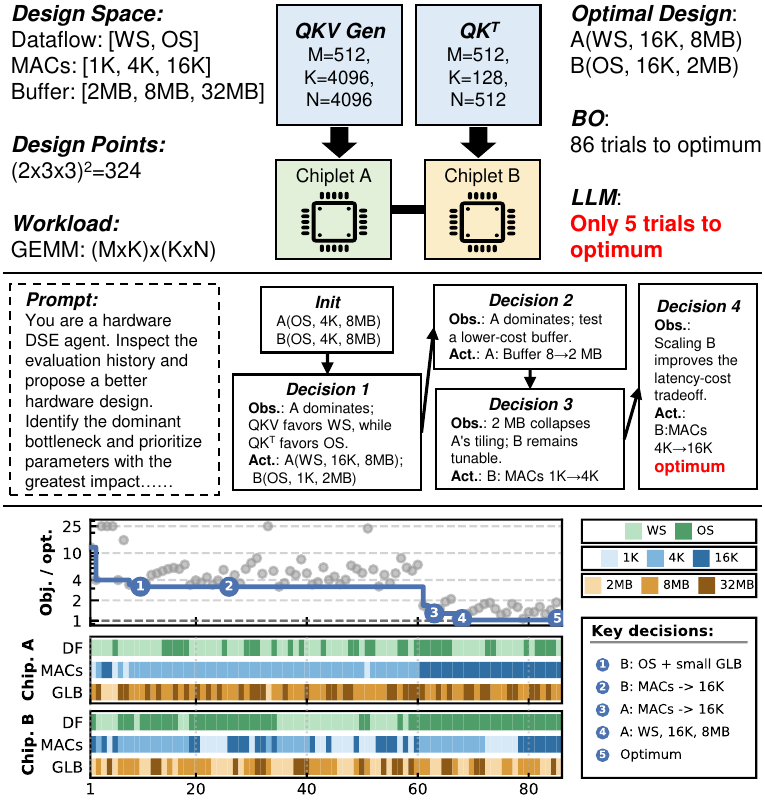}
\caption{A motivating example illustrating the advantage of an LLM over Bayesian optimization (BO). In a search space containing 324 design points, BO requires 86 evaluations to find the optimal design, whereas the LLM requires only five. The middle panel shows the decision process of the LLM. The lower panel shows the BO search trajectory, including the objective normalized to the optimum and the parameter selections at each iteration.}
\label{fig:intro}
\end{figure}

Fig.~\ref{fig:intro} illustrates this difference in information utilization. The case maps QKV generation and QK\(^\top\) computation onto two chiplets and searches for the optimal configuration in a space containing 324 design points. As shown in the lower panel of Fig.~\ref{fig:intro}, BO quickly discovers that QK\(^\top\) favors the output-stationary (OS) dataflow and a small buffer. However, when exploring high-compute configurations for Chiplet A, it combines the weight-stationary (WS) dataflow and 16K MACs with an insufficient 2 MB buffer, causing severe objective degradation. Because BO observes only the final scalar objective, it cannot directly determine whether the failure is caused by the high-compute configuration itself or by insufficient buffer capacity. It therefore remains around 4K MACs for Chiplet A for a long period, and does not discover the effective combination of WS, 16K MACs, and an 8 MB buffer until evaluations 67--68. BO finally reaches the optimum at evaluation 86 after increasing the compute capacity of Chiplet B. In contrast, as shown in the middle panel, the LLM directly identifies QKV generation as the dominant bottleneck from the initial report and selects WS, 16K MACs, and an 8 MB buffer for Chiplet A according to its workload characteristics, while retaining OS and a small buffer for QK\(^\top\). The LLM then observes from subsequent reports that the primary optimization opportunity has shifted to Chiplet B and reaches the optimal design at evaluation 5 by increasing its compute resources.

This case demonstrates that an LLM can extract diagnostic relationships that are not directly available from scalar feedback. Given a poorly performing parameter combination, the LLM can identify the actual cause of degradation, distinguish the effect of an individual parameter from parameter interactions, and determine whether the bottleneck has shifted after a hardware modification. Traditional methods must gradually learn which configurations are generally preferable from multiple samples, whereas an LLM can directly analyze why the current configuration performs poorly and what should be modified next. This diagnostic capability makes each decision more targeted under a limited evaluation budget.

Based on these observations, we propose Beacon, a report-driven LLM multi-agent HW-DSE framework for heterogeneous multi-chiplet accelerators, to improve design decisions under limited evaluation budgets. Beacon organizes each decision into bottleneck localization, root-cause diagnosis, and architecture modification, thereby transforming fine-grained evaluation evidence into targeted and traceable search actions.

The main contributions are summarized as follows:

\begin{enumerate*}[itemjoin=\\\hspace*{\parindent}]
    \item We introduce a report-driven LLM-agent paradigm for HW-DSE of heterogeneous multi-chiplet accelerators. Unlike conventional black-box methods driven mainly by final objective values, it leverages fine-grained evaluator reports to guide bottleneck diagnosis and hardware adjustment under limited evaluation budgets. Different from prior LLM-based chip-design studies focused on EDA flow automation, our work targets architecture-level hardware design space exploration.

    \item We design Beacon, a hierarchical collaborative LLM-agent framework for DSE. Beacon divides the search process into model-level bottleneck discovery, layer-level root-cause diagnosis, and architecture-level candidate generation. It organizes agent evidence acquisition and tool invocation through a ReAct-style \cite{react} interaction scheme. Beacon further integrates a specialized Analysis Toolbox and a RAG \cite{rag} memory module to construct a closed-loop search workflow for HW-DSE. \textbf{Beacon is open-sourced at: \url{https://anonymous.4open.science/r/Beacon-42BC/}}.

    \item We systematically evaluate Beacon on large-model inference workloads. Under the same limited evaluation budget, compared with random search, Bayesian optimization, and reinforcement learning, Beacon reduces the composite latency-energy-monetary-cost objective by 25.1\% to 93.5\%, demonstrating the effectiveness of report-driven search for complex multi-chiplet accelerator DSE.
\end{enumerate*}

\section{Related Work}
\label{sec:related_work}

\subsection{Hardware DSE for Accelerators}
\label{subsec:related_accelerator_dse}

Hardware design space exploration (HW-DSE) for deep learning accelerators can be broadly divided into offline surrogate model methods and online search methods. Offline surrogate model methods collect design samples and train models to predict the quality of candidate hardware configurations. For example, Mind Mappings \cite{mind-mappings}, GANDSE \cite{gandse}, VAESA \cite{vaesa}, and AIRCHITECT \cite{airchitect,airchitectv2} use differentiable surrogate models, generative adversarial networks, variational autoencoders, MLPs, or contrastive learning models to optimize hardware parameters. Online search methods evaluate candidate designs during search and iteratively update the search policy. For example, DiGamma \cite{digamma} and ConfuciuX \cite{confuciux} use genetic algorithms or reinforcement learning to optimize hardware configurations, while HASCO \cite{hasco} and Spotlight \cite{spotlight} use improved Bayesian optimization methods to explore hardware parameters.

These methods provide useful foundations for accelerator parameter search, but their dependence on large scale data sampling makes them difficult to apply directly to heterogeneous multi-chiplet DSE with expensive evaluations.

\subsection{DSE for Multi-Chiplet Accelerators}
\label{subsec:related_chiplet_dse}

Architecture design and DSE for multi-chiplet DNN accelerators mainly study how workloads are mapped and executed across chiplets. Simba \cite{simba} demonstrates a DNN inference accelerator based on a multi-chip module, with hardware design mainly relying on manual architectural configuration. MOHaM \cite{moham} targets multiple DNN workloads and uses a multi-objective genetic algorithm to search Pareto designs. Gemini \cite{gemini} studies joint optimization of architecture and mapping for large scale DNN chiplet accelerators, with hardware candidates mainly enumerated by grid search. Compass \cite{compass} builds a multi-chiplet accelerator evaluator and mapping search framework for LLM inference serving workloads, and applies Bayesian optimization over a predefined hardware space.

These studies provide support for multi-chiplet evaluation and mapping, but their hardware search is usually based on predefined candidates or relatively simple search methods, which limits exploration of the hardware design space.

\subsection{Agents for Chip Design}
\label{subsec:related_agent_chip_design}

LLM agents have recently been applied to chip design. ChatEDA \cite{chateda} uses an LLM as an EDA agent for task decomposition, script generation, and RTL to GDSII flow execution. MAHL \cite{mahl} proposes a hierarchical chiplet design flow guided by multiple LLM agents, with adaptive debugging to support design generation and repair. LayoutCopilot \cite{layoutcopilot} builds an LLM powered collaborative multiagent framework for interactive analog layout design. CROP \cite{crop} combines circuit retrieval, LLM reasoning, and parameter guided optimization for candidate retrieval and improvement in circuit design spaces. Atelier \cite{atelier} uses multiple LLM agents and circuit knowledge bases for automated analog circuit analysis and design decision making.

Compared with the above studies, this work focuses on architecture-level combinatorial optimization of hardware parameters, rather than EDA flow execution or circuit-level optimization. Existing LLM agent works usually center on generating scripts, code, layout operations, or circuit candidates, and mainly rely on feedback from compilation, simulation, rule checking, or human interaction. The core contribution of this work is to show how to design a hierarchical collaboration mechanism for LLM agents in HW-DSE, so that they can use unstructured data for combinatorial optimization. Our work is orthogonal to these studies and fills the gap of applying LLM agents to this setting.

\section{Preliminaries}
\label{sec:preliminaries}

\subsection{Heterogeneous Multi-Chiplet Accelerator}
\label{subsec:prelim_accelerator}

\begin{figure}[ht]
\centering
\includegraphics[width=0.48\textwidth]{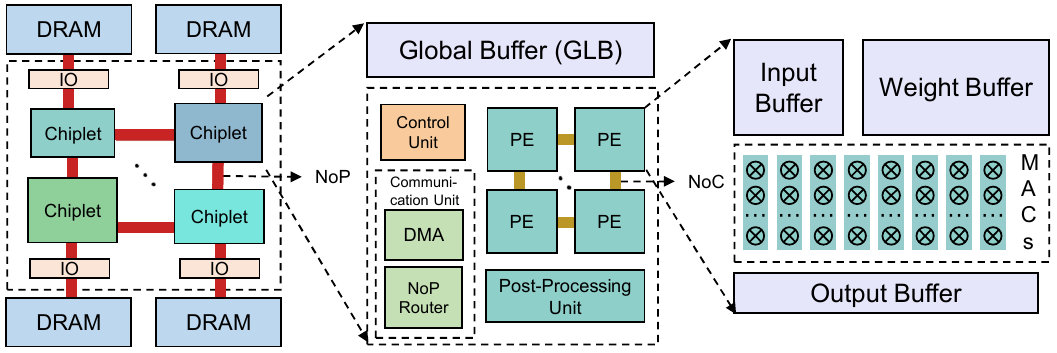}
\caption{Abstraction of the heterogeneous multi-chiplet accelerator.}
\label{fig:background_hardware}
\end{figure}

As shown in Fig.~\ref{fig:background_hardware}, the hardware design considered in this work is based on a general abstraction of multi-chiplet deep learning accelerators. This abstraction is derived from common features in prior chiplet DSE studies \cite{gemini,compass,moham,scar}. By changing configuration parameters, it can represent different multi-chiplet accelerator designs. 

In this abstraction, multiple compute chiplets are connected through a network on package (NoP) to support data exchange across chiplets. Chiplets located at the boundary of the array access external DRAM through IO dies or memory interfaces. The abstraction can represent both homogeneous multi-chiplet accelerators and heterogeneous multi-chiplet accelerators, where different chiplets may use different dataflows or micro-architectures.

Each compute chiplet is similar to a conventional monolithic accelerator. It contains a global buffer (GLB), a processing element (PE) array, a network on chip (NoC), a control unit, and a postprocessing unit. Each PE further contains a parallel multiply accumulate array and more fine grained buffers.

A chiplet is usually assigned the workload of one layer. The layer workload is further tiled and processed by the PE array within the chiplet according to the selected dataflow. Different compute chiplets can have different dataflows and compute or memory configurations, so that they can better match layers with different compute intensity and architectural preferences.

\subsection{Hardware Design Space}
\label{subsec:prelim_design_space}

This work focuses on architecture level hardware design for heterogeneous multi-chiplet accelerators. The design variables include the chiplet count, the configuration of each chiplet, and system parameters shared by the entire accelerator.

First, the search process selects the chiplet count. Let $\mathcal{N}$ denote the set of legal chiplet counts. The selected chiplet count satisfies
\begin{equation}
    N \in \mathcal{N}.
\end{equation}
For each legal chiplet count, the hardware template derives the corresponding two dimensional array shape as
\begin{equation}
    \phi(N) = (H_N, W_N), \qquad H_N W_N = N.
\end{equation}

After determining the chiplet count, the search process independently configures each chiplet in the array. Let $\mathcal{C}$, $\mathcal{B}$, and $\mathcal{D}$ denote the legal sets of compute capacities, on-chip buffer capacities, and dataflow architectures, respectively. The configuration space of one chiplet is
\begin{equation}
    \mathcal{T}
    =
    \mathcal{C}
    \times
    \mathcal{B}
    \times
    \mathcal{D}.
\end{equation}
The compute capacity determines the number of compute units and the PE or MAC organization. The buffer capacity determines the on-chip memory resources, while the dataflow architecture determines the data reuse and execution pattern within the chiplet. The configuration of chiplet $i$ is
\begin{equation}
    t_i = (c_i, b_i, d_i) \in \mathcal{T}.
\end{equation}
Each chiplet can independently select any legal configuration from $\mathcal{T}$. The configuration of the complete chiplet array is
\begin{equation}
    z_{\mathrm{chip}}
    =
    (t_1, t_2, \ldots, t_N)
    \in
    \mathcal{T}^{N}.
\end{equation}

In addition to the chiplet count and per-chiplet configurations, the system parameters shared by the entire accelerator must also be coordinated. We divide them into system-level hardware parameters and system-level mapping parameters. Let $\mathcal{G}_{\mathrm{hw}}$ and $\mathcal{G}_{\mathrm{map}}$ denote their legal spaces, respectively. The system parameter space is
\begin{equation}
    \mathcal{G}
    =
    \mathcal{G}_{\mathrm{hw}}
    \times
    \mathcal{G}_{\mathrm{map}}.
\end{equation}
System-level hardware parameters describe globally shared hardware resources, such as NoP and DRAM bandwidth. System-level mapping parameters describe workload partitioning and execution organization, including tensor parallelism and micro-batch size. Although they are mapping parameters, they closely interact with the hardware configuration and substantially affect execution efficiency. We therefore include them in the search space. A system configuration is represented as
\begin{equation}
    z_{\mathrm{sys}}
    =
    (z_{\mathrm{sys}}^{\mathrm{hw}}, z_{\mathrm{sys}}^{\mathrm{map}}),
    \quad
    z_{\mathrm{sys}}^{\mathrm{hw}} \in \mathcal{G}_{\mathrm{hw}},
    \quad
    z_{\mathrm{sys}}^{\mathrm{map}} \in \mathcal{G}_{\mathrm{map}}.
\end{equation}

A complete design configuration is represented as
\begin{equation}
    z
    =
    (N,
    z_{\mathrm{chip}},
    z_{\mathrm{sys}}).
\end{equation}
The resulting hardware design space is
\begin{equation}
    \mathcal{Z}
    =
    \bigcup_{N \in \mathcal{N}}
    \left(
    \{N\}
    \times
    \mathcal{T}^{N}
    \times
    \mathcal{G}
    \right).
\end{equation}

The search process first selects the chiplet count and derives the array shape. It then independently configures each chiplet and finally selects the system-level hardware and mapping parameters. This design space can represent multi-chiplet accelerator configurations ranging from homogeneous designs to multidimensional heterogeneous designs.

\section{Methodology}
\label{sec:method}

This section first presents the design principles of Beacon. Beacon follows the basic idea of general agentic search methods, but it does not directly apply a generic LLM agent to hardware search. Instead, it is specialized for HW-DSE. Specifically, we organize HW-DSE as a closed-loop process consisting of decision making, memory, and action/observation. The decision capability generates the next search direction according to the current evaluation feedback. The memory capability stores and reuses historical trial-and-error experience. The action/observation capability converts search decisions into evaluable hardware configurations and obtains feedback from the evaluation environment.

Within the decision capability, Beacon further adopts a hierarchical design. This design follows the common process used by hardware architects to analyze performance bottlenecks. Given a hardware evaluation result, the designer usually first observes end-to-end metrics, execution timelines, chiplet runtime states, and execution bubbles to decide which layers or operator groups deserve attention. The designer then analyzes fine-grained compute, memory, and communication behavior in the candidate regions to identify the main bottleneck causes. Finally, the designer chooses hardware parameter updates according to the root causes and hardware constraints. Accordingly, Beacon decomposes the search decision into three consecutive questions: where to optimize, why it is bottlenecked, and how to modify the hardware. It does not rely on a single LLM to read the full report and directly generate a hardware configuration. This hierarchical design reduces the context pressure caused by long reports, lowers the risk of premature parameter modification, and makes each decision step traceable to explicit hardware evidence.

The rest of this section describes the framework implementation in detail.

\begin{figure*}[ht]
\centering
\includegraphics[width=0.95\textwidth]{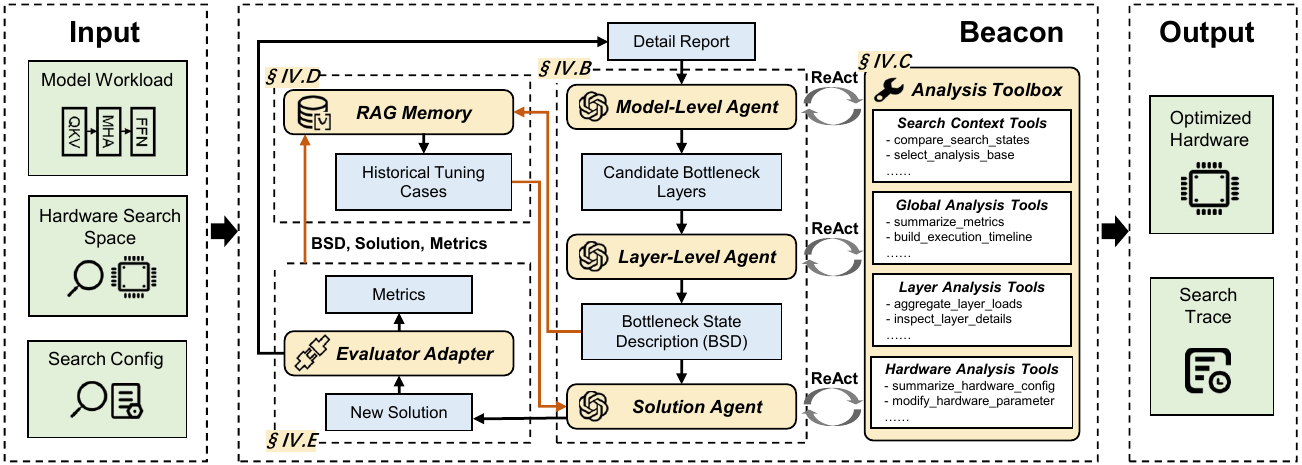}
\caption{Overview of the proposed LLM-based multi-agent hardware search framework.}
\label{fig:framework}
\end{figure*}

\subsection{Overview}
\label{subsec:method_overview}

Fig.~\ref{fig:framework} shows the overall framework. The framework takes three types of inputs: the model workload, the hardware search space, and the search configuration. The model workload specifies the target model and inference scenario, including the layer structure, input size, batch setting, and prefill or decode mode. The hardware search space defines the legal architecture parameters, such as chiplet size, chiplet type, DRAM bandwidth, NoP bandwidth, micro-batch size, and tensor parallelism. The search configuration specifies the initial hardware configuration, objective function, evaluation budget, and output settings. The framework produces two outputs: the optimized hardware configuration found within the evaluation budget, and a search trace that records the metrics, bottleneck states, retrieved cases, solution proposals, tool calls, and hardware modifications in each iteration.

The framework contains four main components. The first component is the \textbf{Agentic Search Engine}, which consists of a Model-Level Agent, a Layer-Level Agent, and a Solution Agent. These agents are responsible for global bottleneck discovery, layer-level bottleneck diagnosis, and hardware candidate generation, respectively. The second component is the \textbf{Analysis Toolbox}. It wraps evaluation reports, search states, and hardware configurations into callable tools that provide structured evidence to the agents. As shown in Fig. \ref{fig:framework}, these tools can be further divided into four categories according to their functional scopes: search context tools, global analysis tools, layer analysis tools, and hardware analysis tools. The third component is \textbf{RAG Memory}. It uses the Bottleneck State Description (BSD) and hardware state as retrieval keys, and returns similar historical search cases as prior experience for solution generation. The fourth component is the \textbf{Evaluator Adapter}. It wraps the underlying accelerator evaluator, converts hardware candidates into standardized metrics and detail reports, and normalizes hardware candidates before evaluation. 

The black arrows in Fig.~\ref{fig:framework} indicate the information flow within one online search iteration. The Evaluator Adapter evaluates the current hardware and produces metrics and detail reports. The agents then use the Analysis Toolbox to collect evidence, proceeding from global bottleneck discovery to layer-level bottleneck diagnosis, and then to hardware candidate generation. The Solution Agent outputs a new solution, which is materialized as a legal hardware candidate and evaluated in the next iteration. The orange arrows indicate the cross-iteration experience flow. The BSD generated by the Layer-Level Agent is used to query RAG Memory. Retrieved historical search cases are provided to the Solution Agent, and each evaluated transition is written back to memory after evaluation. In this way, the framework uses the current evaluation reports for immediate diagnosis while accumulating successful and unsuccessful cases across iterations.

\subsection{Agentic Search Engine}
\label{subsec:agentic_search}

\subsubsection{Prompt and ReAct Design}
\label{subsubsec:prompt_react}

\begin{figure}[ht]
\centering
\includegraphics[width=0.48\textwidth]{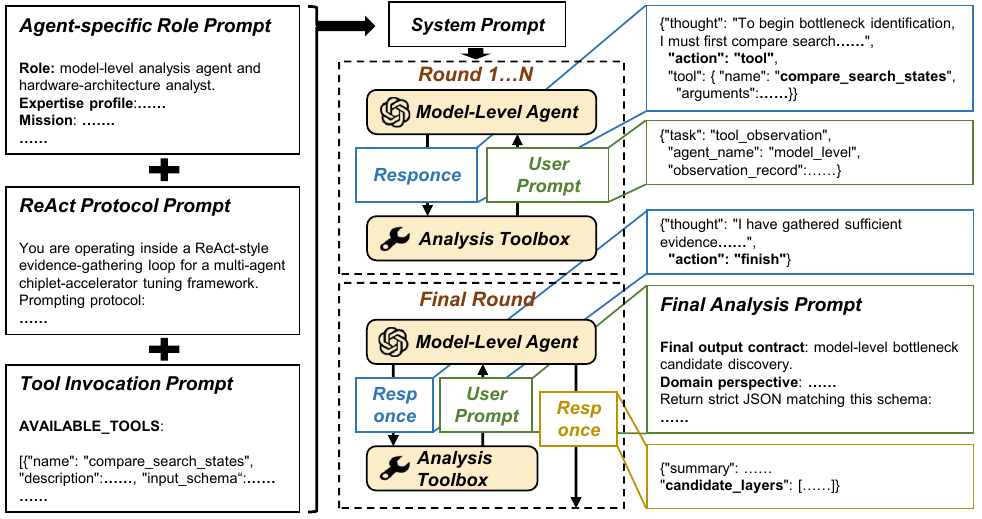}
\caption{The process of the model-level agent executing ReAct interactions.}
\label{fig:model-level-agent}
\end{figure}

The three agents share the same prompt organization, but each agent has a different role prompt, task objective, and output schema. Taking the model-level agent as an example, Fig. \ref{fig:model-level-agent} illustrates the entire process of executing ReAct-style invocations. Each agent invocation has two stages. The first stage is ReAct-style evidence collection, where the agent alternates between LLM decisions and tool calls. The second stage is final synthesis, where the agent produces a structured output based on the collected tool observations.

The system prompt is composed of three parts. The first part is an agent-specific role prompt, which defines the role, task boundary, and expected evidence sources. For example, the Model-Level Agent is defined as an architecture analyst for whole-model execution behavior, the Layer-Level Agent as a layer-level bottleneck diagnosis agent, and the Solution Agent as a hardware design space search agent. The second part is a shared ReAct protocol. It requires each step to return strict JSON and to choose either one tool call or a termination action. The third part is the list of available tools serialized from the Analysis Toolbox, including tool names, descriptions, and argument schemas. Placing tool descriptions in the system prompt makes tool usage less sensitive to later context compression.

In each ReAct step, the agent either requests one tool observation or stops tool use. If a tool is requested, the system executes the tool and appends the structured observation to the conversation. If the agent has enough evidence, or if the step budget is reached, tool use ends. A final output prompt is then appended to request a strict JSON response under the agent-specific schema. The final response is validated before it is passed to downstream components. This design limits the LLM's role to evidence selection, evidence interpretation, and decision generation, while report parsing, hardware normalization, and state storage are handled by deterministic components.

\subsubsection{Model-Level Agent}
\label{subsubsec:model_agent}

The Model-Level Agent identifies candidate bottleneck layers from a whole-model execution perspective. Its inputs include the metrics, detail reports, search state, hardware summary, and toolbox interface for the selected analysis base. The agent first uses Search Context Tools to compare the current, previous, and best evaluated configurations, and then selects the analysis base for the current iteration. This step is necessary because the latest configuration is not always the most informative one. For example, if the previous search move caused a clear regression, analyzing the best configuration can provide a better basis for the next decision.

After the analysis base is selected, the agent uses global analysis tools and layer ranking tools to collect evidence from multiple views. These views include latency, energy, monetary cost, objective changes, inter-layer dependency, core utilization, layer group contribution, and layer rank views. This stage only decides which layers or operator regions require further diagnosis. It does not propose hardware changes.

Table~\ref{tab:model_agent_output} summarizes the structured output of the Model-Level Agent. This design emphasizes global evidence. A layer with high latency is not necessarily the best search target, because it may not be on the critical path or may contribute little to energy or monetary cost. The Model-Level Agent therefore combines multiple global views before selecting candidates for detailed diagnosis.

\begin{table}[t]
    \centering
    \caption{Output fields of the Model-Level Agent.}
    \label{tab:model_agent_output}
    \footnotesize
    \begin{tblr}{
        width = \linewidth,
        colspec = {X[0.32] X[0.58]},
        cells = {halign=c, valign=m},
        row{1} = {font=\bfseries},
        hlines,
        vlines,
    }
        Field & Meaning \\
        Analysis base & Evaluated design used as the analysis base. \\
        Summary & Short summary of whole-model bottlenecks. \\
        Candidate layers & Candidate layers with operator group, concern type, ranking metric, evidence, and confidence. \\
        Bottleneck objective & Main objective of concern, such as latency, energy, monetary cost, or mixed. \\
        Rank views & Layer ranking views used by the agent. \\
        Global findings & Additional whole-model observations. \\
    \end{tblr}
\end{table}

\subsubsection{Layer-Level Agent}
\label{subsubsec:layer_agent}

The Layer-Level Agent receives the candidate layers from the Model-Level Agent and diagnoses why these layers are bottlenecks. Its inputs include the candidate layer list, the detail reports of the selected analysis base, hardware context, search state, and toolbox interface. The agent mainly uses layer analysis tools. It may also use hardware analysis tools when chiplet placement or heterogeneous layout is relevant to communication behavior.

The diagnosis considers compute, memory, and communication dimensions. The compute dimension covers compute-array pressure, MAC utilization, and operator compute density. The memory dimension covers buffer usage and DRAM access behavior. The communication dimension covers NoP transfer overhead. The output of this agent is the Bottleneck State Description (BSD), which is summarized in Table~\ref{tab:bsd_fields}.

\begin{table}[t]
    \centering
    \caption{Main fields in the Bottleneck State Description.}
    \label{tab:bsd_fields}
    \footnotesize
    \begin{tblr}{
        width = \linewidth,
        colspec = {X[0.32] X[0.58]},
        cells = {halign=c, valign=m},
        row{1} = {font=\bfseries},
        hlines,
        vlines,
    }
        Field & Meaning \\
        Primary impact & Main affected objective, such as latency, energy, cost, or mixed. \\
        Dominant root cause & Dominant cause, such as compute, memory, communication, buffer, scheduling, or imbalance. \\
        Layer diagnoses & Per-layer diagnoses with impact types, root causes, ratios, features, and text explanation. \\
        Retrieval description & Compact semantic description used for RAG retrieval. \\
        Root-cause summary & Summary of root-cause dimensions across candidate layers. \\
        Recommended focus & Hardware parameters recommended for the Solution Agent to inspect. \\
    \end{tblr}
\end{table}

BSD is the central intermediate representation in the framework. It compresses complex detail reports into a stable diagnostic state. It also serves as the retrieval key for RAG Memory, so that historical cases are retrieved by bottleneck similarity rather than by hardware JSON similarity alone.

\subsubsection{Solution Agent}
\label{subsubsec:solution_agent}

The Solution Agent converts the BSD, historical search cases, current hardware state, and hardware search-space constraints into the next solution. Its inputs include the BSD from the Layer-Level Agent, retrieved cases from RAG Memory, the current hardware configuration, the search state, and hardware analysis tools.

The agent usually follows three steps. First, it maps the diagnosed bottleneck to candidate hardware parameters. For example, compute or buffer bottlenecks often point to chiplet size, memory bottlenecks may point to DRAM bandwidth or micro-batch size, and communication bottlenecks may point to NoP bandwidth, chiplet type layout, or tensor parallelism. Second, it compares similar historical search cases to identify actions that were effective or ineffective in previous iterations. Third, it uses hardware analysis tools to materialize legal candidates and selects one candidate as the final new solution.

Table~\ref{tab:solution_agent_output} summarizes the output fields of the Solution Agent. The final hardware candidate is not taken directly from free-form LLM output. It is normalized and validated through the hardware tools and the Evaluator Adapter before evaluation.

\begin{table}[t]
    \centering
    \caption{Output fields of the Solution Agent.}
    \label{tab:solution_agent_output}
    \footnotesize
    \begin{tblr}{
        width = \linewidth,
        colspec = {X[0.32] X[0.58]},
        cells = {halign=c, valign=m},
        row{1} = {font=\bfseries},
        hlines,
        vlines,
    }
        Field & Meaning \\
        Strategy & Short name of the search strategy for this iteration. \\
        Hardware update & High-level hardware modification intent. \\
        Selected candidate & Optional reference to a materialized candidate from a tool call. \\
        Actions & Parameter changes, such as bandwidth or chiplet type changes. \\
        Rationale & Reasoning based on bottleneck evidence, history, and cost. \\
        History usage & How retrieved cases influenced the decision. \\
        Cost risk & Expected monetary-cost effect and risk. \\
        Expected objective direction & Expected direction of the overall objective. \\
    \end{tblr}
\end{table}

\subsection{Analysis Toolbox}
\label{subsec:analysis_toolbox}

The Analysis Toolbox provides structured evidence access for the search loop. It follows three design principles. First, the tools should cover the full path from search-state comparison, global metrics, layer diagnosis, to hardware modification. This prevents agents from parsing raw reports directly. Second, the tools should have separated responsibilities. For instance, global metrics and detailed layer evidence are returned by different tools. Third, tool outputs should be compact, structured, and traceable to evaluation files or generated artifacts.

Table~\ref{tab:toolbox} lists the current tools. The four categories correspond to the groups shown in Fig.~\ref{fig:framework}. Search context tools select the analysis target. Global analysis tools provide whole-model evidence. Layer analysis tools support layer-level diagnosis. Hardware analysis tools provide hardware-state summaries and legal candidate editing.

\begin{table*}[t]
    \centering
    \caption{Tools in the Analysis Toolbox.}
    \label{tab:toolbox}
    \footnotesize
    \providecommand{\ToolCell}[2]{\begin{tabular}[c]{@{}c@{}}\texttt{#1}\\\texttt{#2}\end{tabular}}
    \begin{tblr}{
        width = \textwidth,
        colspec = {X[0.14] X[0.21] X[0.51] X[0.12]},
        cells = {halign=c, valign=m},
        row{1} = {font=\bfseries},
        hlines,
        vlines,
    }
        Category & Tool & Function & Typical user \\
        \SetCell[r=2]{c} Search Context & \ToolCell{compare\_}{search\_states} & Compares current, previous, and best evaluated designs, and summarizes metric changes, hardware changes, and search trends. & All agents \\
        & \ToolCell{select\_}{analysis\_base} & Selects the active evaluated design used by subsequent analysis tools in the current iteration. & All agents \\
        \SetCell[r=4]{c} Global Analysis & \texttt{summarize\_metrics} & Summarizes latency, energy, monetary cost, and the derived objective for an evaluated design. & Model-Level; Solution \\
        & \ToolCell{build\_execution\_}{timeline} & Builds an execution timeline and reports critical-path behavior, core utilization, and idle intervals. & Model-Level \\
        & \ToolCell{summarize\_}{operator\_groups} & Aggregates latency, energy, and occurrence statistics by operator group, such as attention, projection, and FFN. & Model-Level; Layer-Level \\
        & \ToolCell{summarize\_}{monetary\_cost} & Reports total monetary cost, component costs, and component cost shares. & Model-Level; Solution \\
        \SetCell[r=3]{c} Layer Analysis & \ToolCell{summarize\_layer\_}{rank\_views} & Ranks layers from views such as latency, energy, ending time, compute, memory, communication, and buffer pressure. & Model-Level \\
        & \ToolCell{aggregate\_}{layer\_loads} & Aggregates per-layer load information and returns compact views by summary, ranked layers, or operator groups. & Model-Level; Layer-Level \\
        & \ToolCell{inspect\_}{layer\_details} & Inspects specified layers and returns timing, energy, placement, compute, memory, communication, and buffer evidence. & Layer-Level \\
        \SetCell[r=3]{c} Hardware Analysis & \ToolCell{summarize\_}{hardware\_config} & Summarizes chiplet count, layout, types, compute and memory resources, system parameters, and search-space position. & All agents \\
        & \ToolCell{modify\_}{hardware\_parameter} & Sets one or more legal hardware parameters to target values and returns a materialized candidate. & Solution \\
        & \ToolCell{step\_}{hardware\_parameter} & Applies a one-step legal neighborhood move to chiplet size, chiplet type, or system parameters. & Solution \\
    \end{tblr}
\end{table*}

Search context tools answer which evaluated design should be analyzed. This is needed because the current design, previous design, best design, and recent trajectory may all be relevant. Global analysis tools answer where the main whole-model issues are. They convert scalar metrics, timelines, operator groups, and cost breakdowns into evidence for candidate selection. Layer analysis tools answer why specific layers are bottlenecks. They further explain candidate layers in terms of compute, memory, or communication behavior. Hardware analysis tools answer how the search can move within the legal hardware space. The resulting evidence chain progresses from coarse observations to concrete hardware actions, and each tool call is recorded in the agent trace.

\subsection{RAG Memory}
\label{subsec:rag_memory}

The framework uses Retrieval-Augmented Generation (RAG) \cite{rag} to reuse historical search experience. RAG Memory stores and retrieves historical search cases, and injects similar cases into the Solution Agent context. The final hardware candidate is still generated by the Solution Agent using the current BSD, retrieved cases, and hardware search space.

RAG Memory organizes cases as records in a vector database. After each candidate hardware evaluation, the framework combines the current BSD, the hardware configurations before and after the action, the action proposed by the Solution Agent, and the measured metric feedback to construct a Historical Search Case. Each case is stored as a key-value record. The key is used for similarity retrieval, while the value stores the full search feedback.

The key has two channels. The bottleneck key is derived from the retrieval description, the structured BSD, and a small amount of context. It describes the semantic form of the bottleneck. The hardware key is derived from the normalized hardware configuration. It describes the hardware setting in which the case occurred. A query is built in the same way from the current BSD and current hardware state. This design makes retrieval depend on both problem similarity and hardware-context similarity.

Formally, the bottleneck key and hardware key are embedded as
\begin{equation}
    v_b = \mathrm{Emb}(\mathrm{BSD}), \quad
    v_h = \mathrm{Emb}(\mathrm{Hardware}).
\end{equation}
They are concatenated into a joint retrieval vector,
\begin{equation}
    v = [v_b; v_h].
\end{equation}
The bottleneck channel matches search experience under similar root causes, while the hardware channel reduces direct transfer from cases with very different hardware configurations. Retrieval is performed using cosine similarity between the query vector and case vectors. The top-$k$ Historical Search Cases are then returned to the Solution Agent. As iterations proceed, Memory accumulates both successful and unsuccessful cases, so later solution generation can use evaluated search experience.

\subsection{Evaluator Adapter}
\label{subsec:evaluator_adapter}

The Evaluator Adapter is the boundary layer between Beacon and the underlying accelerator evaluator. It isolates evaluator-specific hardware formats, execution scripts, and report organizations, so that the upper-level Agentic Search Engine and Analysis Toolbox only operate on a unified evaluation abstraction. A new evaluator can be connected to the framework as long as it accepts the same form of model workload and hardware candidate, and returns standardized metrics and fine-grained reports.

For one evaluation, the adapter receives two types of inputs: the model workload and the hardware candidate. The former describes workload information such as the model structure, request organization, and inference stage. The latter specifies the concrete hardware configuration generated by the current search step. After invoking the underlying evaluator, the adapter converts the result into unified metrics and detail report. Metrics include latency, energy, monetary cost, and the derived objective. The detail report compactly preserves fine-grained evidence for later analysis, including per-layer or per-core execution intervals, compute/NoP/DRAM time components, energy breakdowns, hardware cost breakdowns, and the evaluated hardware snapshot. Based on these standardized results, the Analysis Toolbox further constructs timelines, layer-level load summaries, operator-group statistics, and hardware summaries.

The adapter also normalizes hardware candidates before they enter the evaluator. The new solution generated by the Solution Agent is often a high-level hardware modification intent rather than a complete evaluator configuration. The adapter materializes it into a legal candidate under the hardware search space, fills in derived fields, and checks duplicate configurations before invoking the evaluator. In this way, the LLM only proposes the search direction, while hardware legality and evaluation file organization are guaranteed by deterministic procedures.

\section{Experimental Evaluation}
\label{sec:experiments}

\subsection{Experimental Setup}
\label{subsec:exp_setup}

\subsubsection{DSE Setup}
\label{subsubsec:dse_setup}

The main experiments cover three model sizes and their corresponding hardware compute scales. We design multi-chiplet accelerators with 64, 512, and 2048 TOPS for GPT3-7B, GPT3-13B \cite{gpt3}, and LLaMA3-70B \cite{llama,llama3}, respectively. GPT3 follows the conventional Transformer architecture. LLaMA3 adopts more recent architectural features, such as pre-layer normalization \cite{pre_layer_norm} and grouped-query attention (GQA) \cite{gqa}. The hardware compute scales are matched to the model sizes. They are also aligned with the compute ranges of Simba \cite{simba}, A100 \cite{a100}, and H100 \cite{h100}, respectively, covering edge-side, medium-scale cloud, and high-end cloud inference accelerators.

The design objective is to minimize
\begin{equation}
    \mathcal{O}(z)
    =
    L(z) \cdot E(z) \cdot MC(z)
    \cdot
    \frac{\max(P(z), P_t)}{\min(P(z), P_t)},
    \label{eq:exp_objective}
\end{equation}
where $L(z)$ denotes latency, $E(z)$ denotes energy, and $MC(z)$ denotes monetary cost. $P(z)$ and $P_t$ denote the actual compute capability of candidate hardware $z$ and the target compute capability, respectively. Instead of using a hard constraint on the exact compute budget, we convert the target compute requirement into a soft penalty. This avoids wasting many search steps when Bayesian optimization or reinforcement learning generates candidates that do not exactly match the target compute budget. It also provides flexibility in the design process. The penalty gives symmetric multiplicative punishment to both over-provisioned and under-provisioned configurations.

\subsubsection{Evaluation Framework}
\label{subsubsec:evaluation_framework}

We use a modified GEMINI \cite{gemini} framework as the simulator. GEMINI is a representative simulator for DSE of multi-chiplet accelerators. We extend it to support LLM inference. We use the genetic algorithm adopted in Compass \cite{compass} to search mapping solutions. The population size and the number of generations are kept the same as in the original setting, namely 120 and 100.

The GovReport \cite{govreport} dataset is used to provide the sequence length distribution. It is a commonly used dataset in long-context inference studies. We follow the chunked prefill paradigm used in recent inference engines to organize requests. Following the default setting of vLLM \cite{vllm}, the chunk size and batch size are set to 2048 and 128, respectively. Each batch contains one chunked prefill request and 127 decode requests.

\subsubsection{Hardware Parameters}
\label{subsubsec:hardware_parameters}

We use the same DSE-irrelevant hardware parameters as GEMINI. The process technology is TSMC 12 nm. The package substrate is an organic substrate. The clock frequency is set to 1 GHz. The NoP adopts GRS technology, a mesh topology, and XY routing. The accelerator is equipped with four DRAM chips, with two placed on each side of the chiplet array.

Table~\ref{table:exp:hardware_range} lists the hardware configuration candidates related to DSE. The relationships among these parameters follow the definition in Section~\ref{sec:preliminaries}.

\begin{table}[t]
    \centering
    \caption{The range of hardware parameters for DSE}
    \label{table:exp:hardware_range}
    \footnotesize
    \begin{tblr}{
        width = \linewidth,
        colspec = {Q[95] Q[370] Q[520]},
        cells = {halign=c, valign=m},
        hlines,
        vlines,
    }
        $N$ & Number of Chiplets & $[1,2,4,8,16,32,64,128,256]$ \\
        \SetCell[r=3]{c} $z_{\mathrm{chip}}$ & Chiplet Dataflow & Weight Stationary (WS), Output Stationary (OS) \\
        & Chiplet MACs & $[4,8,16,32,64,128,256,512]$ K \\
        & Chiplet GLB Size & $[2,4,8,16,32,64]$ MiB \\
        \SetCell[r=2]{c} $z_{\mathrm{sys}}^{\mathrm{hw}}$ & NoP Bandwidth & $[32,64,128,256,512]$ GB/s \\
        & DRAM Bandwidth per Chip & $[16,32,64,128,256]$ GB/s \\
        \SetCell[r=2]{c} $z_{\mathrm{sys}}^{\mathrm{map}}$ & Tensor Parallelism & $[1,2,4,8,16,32,64]$ \\
        & Micro-Batch Size & $[1,2,4,8,16,32,64,128]$ \\
    \end{tblr}
\end{table}

\subsubsection{Baseline Setup}
\label{subsubsec:baseline_setup}

As discussed in Section~\ref{sec:introduction}, existing methods for HW-DSE can be broadly classified into offline and online approaches. Offline methods generally require substantial training data and are unsuitable for our evaluation-budget-limited setting. Bayesian optimization (BO) \cite{compass} and reinforcement learning (RL) \cite{chiplet-gym,confuciux} are the most representative online approaches. We therefore adapt SOTA techniques from these two lines of work as comparison baselines.

We select random search, BO, and RL as the three baselines. For a fair comparison, we define a unified action space. This action space is comparable to the hardware editing capability provided by the Beacon toolbox, and it ensures that all design points are reachable. The three baselines use their own policies to choose the action at each step.

The unified action space is organized according to the three groups of design variables in Table \ref{table:exp:hardware_range}:
\begin{enumerate}
    \item $N$: Increase or decrease the chiplet count to the adjacent legal value.
    \item $z_{\mathrm{chip}}$: Increase or decrease the MAC count or GLB size of one chiplet or all chiplets to the adjacent legal value, or change their dataflow type.
    \item $z_{\mathrm{sys}}$: Increase or decrease a system-level hardware parameter, including NoP and DRAM bandwidth, or a system-level mapping parameter, including tensor parallelism and micro-batch size, to the adjacent legal value.
\end{enumerate}

The implementation of each baseline is described as follows.

\textbf{Random search.}
At each step, one action is randomly selected from the action space and then executed.

\textbf{Bayesian optimization (BO).}
Following Compass~\cite{compass}, BO uses a Gaussian process as the surrogate model and expected improvement (EI) as the acquisition function. The kernel is a composite kernel. It uses a standard radial basis function (RBF) kernel to measure distances among system parameters. When two designs have the same chiplet array shape, it further incorporates the topology-aware layout similarity term proposed in Compass. At each iteration, BO generates a set of candidate hardware configurations from the unified action space and selects the next evaluation point according to EI.

\textbf{Reinforcement learning (RL).}
Following Chiplet-Gym \cite{chiplet-gym}, we construct the state representation by encoding hardware resources and performance feedback in a structured form. Inspired by ConfuciuX \cite{confuciux}, we then formulate the search as a sequential decision process. At each iteration, the hardware configuration evaluated in the previous iteration serves as the current optimization target, and the policy selects one legal modification from the unified action space. The modified hardware configuration is subsequently submitted for evaluation. The reward is defined as
\begin{equation}
r = \log \frac{\mathcal{C}_{\mathrm{prev}}}{\mathcal{C}_{\mathrm{new}}},
\label{eq:exp_rl_reward}
\end{equation}
where $\mathcal{C}_{\mathrm{prev}}$ denotes the objective value before the action, and $\mathcal{C}_{\mathrm{new}}$ denotes the objective value after the action. The policy model adopts an actor-critic policy-gradient method and is updated online using the clipped policy objective commonly employed in PPO.

\subsubsection{Experimental Environment and Hyperparameters}
\label{subsubsec:environment_hyperparameters}

We use DeepSeek-V4-Pro as the LLM backend of Beacon. The model is open-weight, has strong capability among current LLMs, and has relatively low API cost, making it suitable for practical use.

RAG Memory uses BGE-small-en-v1.5 \cite{embedding-model} as the embedding model. This is a commonly used embedding model. Retrieval returns the top-5 historical cases. This setting is determined by preliminary experiments as a tradeoff between search effectiveness and context length.

The experiments are conducted on a server with two Hygon 7285 processors, providing 128 logical cores. The server is also equipped with four A100 GPUs. The simulator and the baselines mainly run on CPUs. GPUs are mainly used for embedding inference and RL model updating. All methods start from the same initial hardware and perform 100 rounds of hardware search. This setting reflects the performance of different methods under a limited evaluation budget.

\subsection{Comparison with Baselines}
\label{subsec:baseline_comparison}

\begin{figure*}[t]
    \centering
    \includegraphics[width=0.9\textwidth]{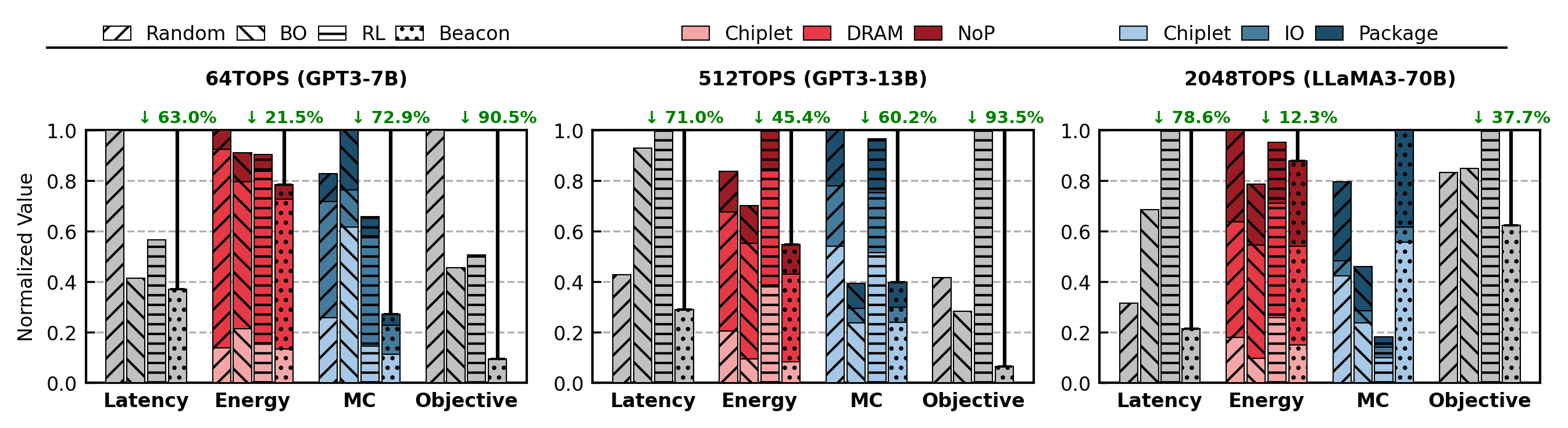}
    \caption{Main comparison across the three compute scales. Each group reports normalized latency, energy, monetary cost (MC), and the composite objective. Energy and MC are further decomposed into their major components. Lower values indicate better designs.}
    \label{fig:main_results}
\end{figure*}

Fig.~\ref{fig:main_results} compares latency, energy, MC, and the objective after normalization within each metric group. The energy and MC bars are stacked to show their component breakdowns. Overall, Beacon outperforms random search, BO, and RL across all three compute scales. Compared with the strongest traditional baseline in each setting, Beacon reduces the objective by 79.1\%, 77.0\%, and 25.1\% on the 64TOPS, 512TOPS, and 2048TOPS tasks, respectively. This shows that LLM-agent-guided search can substantially improve search efficiency under a limited evaluation budget.

The gap is more pronounced in the 64TOPS and 512TOPS settings. For 64TOPS, the objectives of random search, BO, and RL are $10.51{\times}$, $4.79{\times}$, and $5.31{\times}$ that of Beacon, respectively. For 512TOPS, the three baselines are $6.40{\times}$, $4.35{\times}$, and $15.34{\times}$ that of Beacon. These traditional methods can only infer the coupled effects of hardware changes from the final scalar objective. Therefore, under a limited evaluation budget, they tend to select configurations that are locally beneficial but have a high overall cost. In the 2048TOPS setting, Beacon has a smaller relative advantage, but it still achieves the best objective among all methods. This setting involves a larger model and higher parallelism. Candidate designs with more chiplets can significantly reduce latency, but they also introduce higher monetary cost. Beacon achieves a more stable tradeoff among latency, energy, and MC.

The component breakdowns reveal a scale-dependent trend.
At 64 TOPS, DRAM access dominates energy at 75.5\%, while
chiplet and I/O costs contribute comparable shares of MC
(42.1\% and 41.9\%). At 512 TOPS, the DRAM energy share
decreases to 63.3\% and the NoP share increases to 21.3\%;
meanwhile, chiplet and package costs rise to 60.4\% and 24.9\%
of MC, respectively. At 2048 TOPS, the DRAM and NoP energy
shares become comparable at 44.5\% and 38.4\%, while package
cost reaches 38.4\% and I/O cost falls to 5.9\%. As the compute scale grows, the dominant overhead gradually shifts
from external memory and I/O toward inter-chiplet communication,
chiplets, and packaging.

\subsection{Result Analysis}
\label{subsec:result_analysis}

\begin{table}[t]
    \centering
    \caption{Best hardware configurations found by Beacon.}
    \label{tab:best_hardware}
    \scriptsize
    \begin{tblr}{
        width = \columnwidth,
        colspec = {Q[c,m,wd=0.20\columnwidth] X[c,m] X[c,m] X[c,m]},
        cells = {halign=c, valign=m},
        column{1} = {font=\bfseries},
        row{1} = {font=\bfseries},
        hlines,
        vlines,
    }
        Scale & 64 TOPS & 512 TOPS & 2048 TOPS \\
        Chiplets & 2 (1$\times$2) & 8 (2$\times$4) & 32 (4$\times$8) \\
        Dataflow & WS$\times$1, OS$\times$1 & WS$\times$7, OS$\times$1 & WS$\times$25, OS$\times$7 \\
        MACs & 16K$\times$2 & 32K$\times$8 & 32K$\times$32 \\
        GLB & 8 MiB$\times$2 & 8 MiB$\times$8 & \SetCell{c}{16 MiB$\times$31,\\32 MiB$\times$1} \\
        DRAM/NoP & 16/32 & 16/32 & 64/128 \\
        TP & 8 & 8 & 16 \\
        MB size & 64 & 32 & 32 \\
    \end{tblr}
\end{table}

Table~\ref{tab:best_hardware} summarizes the best hardware found by Beacon. The designs do not simply scale all resources linearly with the target compute capability. Instead, Beacon trades off chiplet count, dataflow, bandwidth, and execution parameters according to workload behavior and cost feedback.

For 64TOPS, Beacon selects a two-chiplet design with one WS chiplet and one OS chiplet. The selected micro-batch size is 64. This configuration has an MC of only \$15.55, which is much lower than the high-bandwidth or larger-chiplet-count configurations commonly found by the baselines. For a small compute target, adding more chiplets or bandwidth quickly increases MC, while the latency benefit is limited. Therefore, a better strategy is to keep only a small number of chiplets, use heterogeneity to cover different operator preferences, and keep DRAM/NoP bandwidth cost low.

For 512TOPS, Beacon finds a lightweight heterogeneous design with eight chiplets, including seven WS chiplets and one OS chiplet. This design preserves enough WS chiplets for weight-reuse-intensive GEMMs, such as QKV projection and FFN. It also introduces a small number of OS chiplets to improve part of the attention-related computation. The result shows that heterogeneity does not mean using as many OS chiplets as possible. Instead, it should be configured according to bottleneck layers and cost feedback. Too many OS chiplets can reduce the weight-reuse benefit of dominant GEMMs and introduce load imbalance.

For 2048TOPS, Beacon uses 32 chiplets, including 25 WS chiplets and 7 OS chiplets. This configuration uses a larger chiplet count to reduce latency, which is suitable for a high-throughput setting where latency still has a large impact on the objective. At the same time, its MC reaches \$593.69, indicating that chiplet count and package communication cost become major constraints at high compute scales. This result also shows that the design challenge in high-end configurations is not simply to increase parallelism. It is necessary to balance parallelism, communication cost, and dataflow heterogeneity.

\begin{figure}[t]
    \centering
    \includegraphics[width=0.49\textwidth]{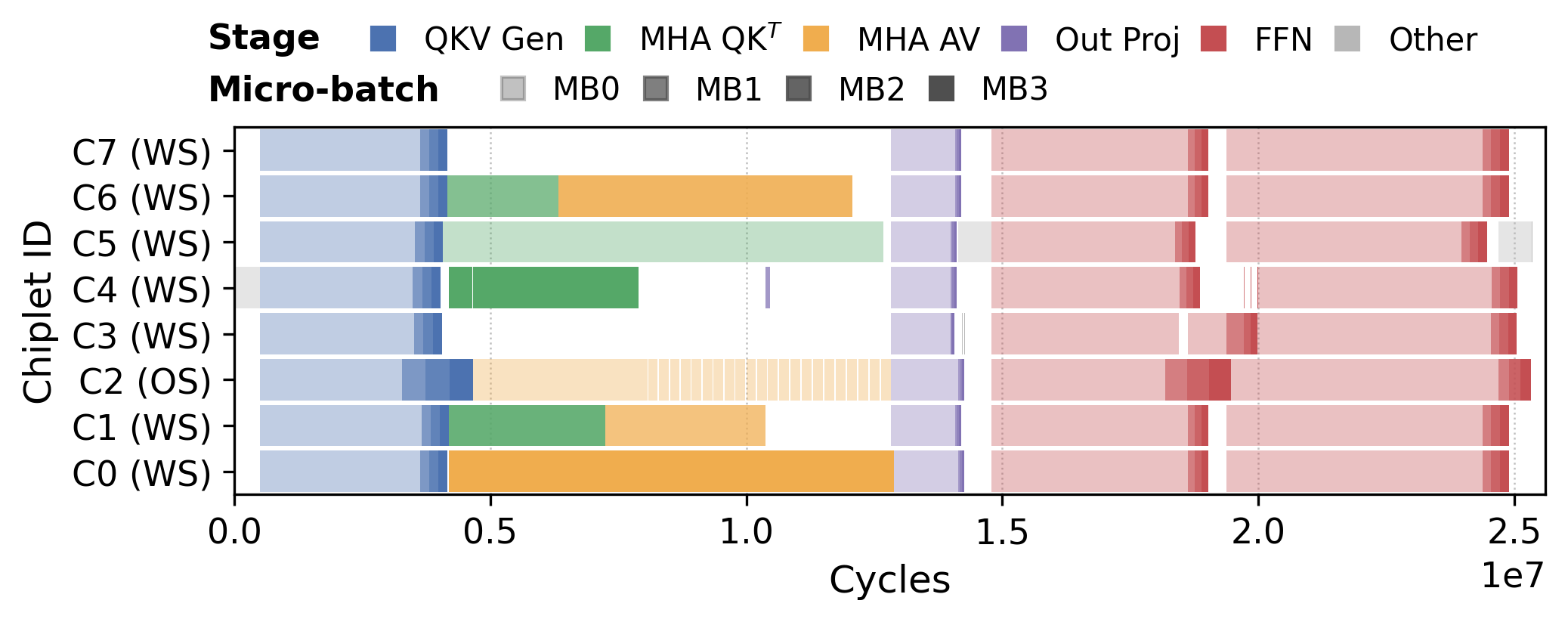}
    \caption{Execution timeline of the 512TOPS best design. Colors represent QKV Gen, MHA QK$^T$, MHA AV, Out Proj, FFN, and Other stages, while color intensity distinguishes micro-batches.}
    \label{fig:timeline}
\end{figure}

Fig.~\ref{fig:timeline} shows the spatiotemporal execution schedule of the 512TOPS best design. This design selects eight chiplets and a tensor-parallel degree of 8. Therefore, tensor-parallel stages such as QKV Gen, Out Proj, and FFN can be evenly partitioned across chiplets, reducing idle resources. This also demonstrates that system parameters must be co-designed to execute the workload efficiently. The batch is divided into four micro-batches. MB0 contains the chunked prefill request, while the other micro-batches correspond to decode requests. As a result, MB0 has longer execution fragments across stages.

The timeline shows a pipeline-like execution pattern. Different stages are assigned to different chiplets, and multiple micro-batches flow through these stages in sequence. This allows each chiplet to reuse resident weights and reduces the latency and energy overhead caused by repeatedly loading weights. The dataflow assignment further shows that the only OS chiplet, C2, handles the long MHA AV computation in MB0. For MB0, the AV stage multiplies the attention-score matrix with the value matrix. In this GEMM, the input matrix is larger than the weight matrix. For example, for one prefill AV head, the attention-score matrix is $2048{\times}2048$, and the value matrix is $2048{\times}128$. This shape is more suitable for the OS dataflow. This result illustrates how a heterogeneous design provides flexible support for different workload stages.

\subsection{Design Process Analysis}
\label{subsec:design_process_analysis}

\begin{figure}[t]
    \centering
    \includegraphics[width=0.49\textwidth]{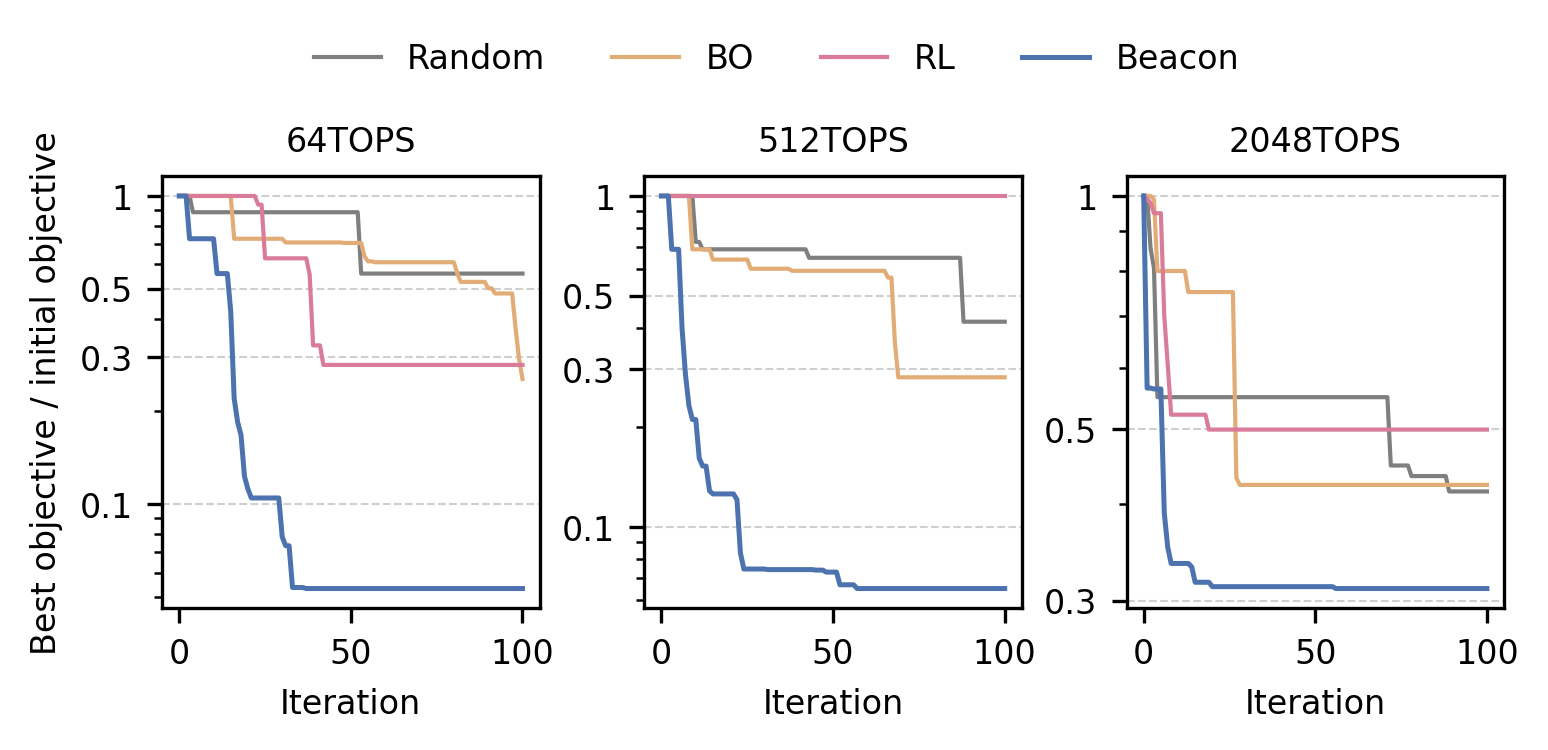}
    \caption{Search curves on the three main tasks. The y-axis reports the objective normalized to the initial design. Lower values indicate better designs.}
    \label{fig:search_curves}
\end{figure}

Fig.~\ref{fig:search_curves} shows the search curves of the three main tasks. Beacon usually reduces the objective rapidly within the first dozen iterations. It then enters a stage of finer-grained cost and heterogeneous-layout adjustment. In contrast, random search lacks direction. BO can fit trends from historical scalar objectives, but it cannot directly identify whether a regression comes from bandwidth cost, dataflow mismatch, or an unsuitable micro-batch setting. RL also has difficulty learning a stable policy within the limited evaluation budget.

For 64TOPS, the search process first explores whether the hardware is over-provisioned. In iteration 3, Beacon tries to reduce the two-chiplet design to a single chiplet. This lowers MC but significantly worsens latency, showing that even the small-scale setting still needs at least two chiplets to support pipeline parallelism. In iteration 11, Beacon restores the two-chiplet design and calibrates per-chiplet MACs to better match the target compute constraint. The following iterations show that the bottleneck is no longer chiplet count. Instead, the execution granularity and sharding granularity are mismatched. TP=16 is too fine for a two-chiplet configuration, and a very small micro\_batch\_size increases scheduling and data-movement overhead. Therefore, iterations 15 to 21 increase micro\_batch\_size and reduce tensor parallelism from 16 to 8, mainly reducing latency and energy. After confirming that bandwidth is not the current bottleneck, iterations 30 to 33 gradually reduce NoP and DRAM bandwidth, mainly reducing MC. Finally, iteration 37 reintroduces one OS chiplet according to the behavior of attention layers, improving part of the attention computation without increasing cost.

For 512TOPS, the search process better reflects the value of hierarchical diagnosis. Early exploration shows that the initial mixed dataflow layout is not suitable for the dominant GEMMs. Iteration 3 switches to eight WS chiplets and reduces energy. Iterations 6 to 12 then increase micro\_batch\_size and reduce tensor parallelism, aligning tensor-parallel shards with the chiplet count and reducing latency and energy. In the middle stage, the model-level analysis identifies the critical tail around FFN2, while the layer-level analysis shows that MHA QK$^T$/AV contributes high energy and latency and has clear chiplet-utilization imbalance. Therefore, iterations 14 to 15 introduce OS chiplets and explore their placement. Iterations 22 to 24 further reduce DRAM and NoP bandwidth according to cost feedback, trading a small latency increase for a significant MC reduction. Later OS-placement trials show that too many OS chiplets introduce load imbalance. Thus, iteration 57 reduces the OS count to one. The final design lets this OS chiplet serve the OS-suitable attention AV fragment while keeping most chiplets as WS for QKV and FFN stages.

For 2048TOPS, the search process follows a different pattern. The initial improvement comes from adjusting micro\_batch\_size. In iterations 1, 6, 7, and 8, Beacon increases micro\_batch\_size from 1 to 16. This reduces latency from $7.3{\times}10^7$ to $2.2{\times}10^7$ and reduces the objective by 66.5\% relative to the initial design. Later results show that increasing the value further to 32 or 64 is not always effective. The setting must match the chiplet count and tensor-parallel degree. Iterations 3, 14, and 20 repeatedly convert low-utilization OS chiplets to WS, reflecting that QKV and FFN GEMMs with large weights remain dominant at this scale. Beacon then tries increasing tensor parallelism, increasing bandwidth, enlarging buffers, expanding to 64 chiplets, and shrinking to 16 chiplets. These actions usually introduce energy, MC, or compute-scale penalties and do not become new best designs. Finally, iteration 56 uses the diagnosis of attention energy and the memory-bound FFN2 tail to add local OS dataflow and a larger buffer at a key position, producing a small improvement. This process shows that the main challenge at high compute scale is not to keep increasing parallelism. Instead, Beacon must identify low-utilization resources and control the side effects of heterogeneity and cost.

\subsection{LLM Backend Selection}
\label{subsec:llm_backend_selection}

\begin{figure}[t]
    \centering
    \includegraphics[width=0.49\textwidth]{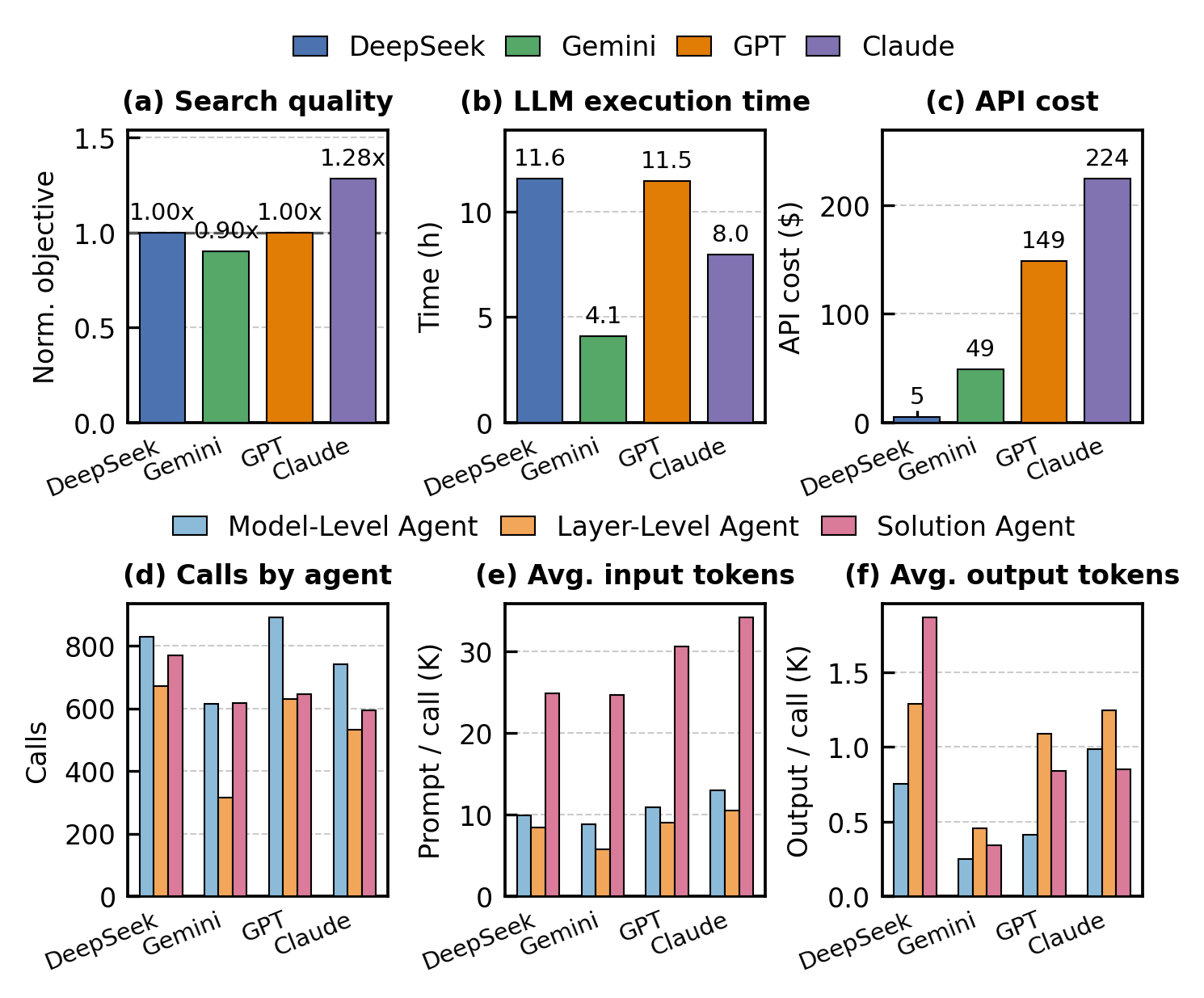}
    \caption{Comparison of LLM backends on the 512TOPS task. The top row reports search quality, total LLM API time, and estimated API cost. The bottom row reports the number of calls, average input tokens per call, and average output tokens per call for the three agents.}
    \label{fig:llm_backend}
\end{figure}

We use the 512TOPS task as the representative setting for LLM-backend analysis. Its compute scale lies between the edge-side and high-end cloud settings, making it representative for both hardware and workload behavior. We compare the default DeepSeek backend with three mainstream LLMs, namely Gemini-3.1-Pro, GPT-5.5, and Claude-Opus-4.8. Fig.~\ref{fig:llm_backend} reports the search quality and cost of these backends.

From the search-quality perspective, all four backends find designs that are significantly better than the initial design, but their optimization preferences differ. Gemini achieves the lowest objective, which is $0.90{\times}$ that of DeepSeek. It selects a four-chiplet design and reduces MC from \$48.71 to \$30.26, but its latency increases from $2.54{\times}10^7$ to $3.77{\times}10^7$. GPT is almost tied with DeepSeek. Its objective is only 0.03\% higher, and both models converge to a region with low bandwidth, TP=8, and MB=32. Their difference lies in dataflow assignment. GPT selects six WS chiplets and two OS chiplets, while DeepSeek selects seven WS chiplets and one OS chiplet. Claude also selects an eight-chiplet heterogeneous design with six WS chiplets and two OS chiplets, but it keeps higher DRAM/NoP bandwidth. This design has lower latency, but its MC rises to \$70.97. As a result, its final objective is $1.28{\times}$ that of DeepSeek.

From the calling-time perspective, Gemini has the shortest total API time, 4.09 hours. Claude takes 7.97 hours, while DeepSeek and GPT take 11.56 and 11.45 hours, respectively. This difference shows that LLM calling time is not directly correlated with final search quality. Gemini is the fastest backend and also finds the best objective in this task. When the calls are broken down by agent, the distribution is similar across backends, and the Model-Level Agent usually accounts for the largest number of calls.

From the token-usage perspective, input tokens dominate the total token count. The Solution Agent has clearly higher average input tokens than the other two agents. This is because it needs to read retrieved RAG cases. The four backends show similar input-token trends, but their output preferences differ. DeepSeek tends to produce longer responses, especially in the Solution Agent stage, while Gemini gives more concise outputs. Consequently, Gemini uses 23.4M total tokens, lower than the 35.9M, 36.8M, and 37.3M tokens used by DeepSeek, GPT, and Claude, respectively.

From the API-cost perspective, we estimate cost using the measured token usage and the official API prices of each provider. DeepSeek costs only \$4.94, which is much lower than Gemini (\$48.98), GPT (\$149.03), and Claude (\$224.47). Overall, different LLM backends affect the search trajectory, but their best designs are broadly comparable and are much better than non-LLM baselines. This supports the robustness of the proposed method. DeepSeek is the most cost-efficient backend, while Gemini has better search quality and shorter wall-clock time in this task. Both are practical choices for Beacon under different cost and runtime priorities.

\subsection{Ablation Study}
\label{subsec:ablation_study}

\begin{table}[t]
    \centering
    \caption{Ablation study on the 512 TOPS task. Norm. is computed by normalizing each objective to that of Beacon.}
    \label{tab:ablation}
    \scriptsize
    \newcommand{\abthead}[1]{%
    \begin{tabular}[c]{@{}c@{}}#1\end{tabular}%
}

\begin{tblr}{
    width = \columnwidth,
    colspec = {X[2.0,c,m] X[0.8,c,m] X[0.8,c,m] X[0.8,c,m] X[0.8,c,m] X[0.9,c,m] X[0.8,c,m]},
    cells = {halign=c, valign=m},
    colsep = 2pt,
    hlines,
    vlines,
    row{1} = {font=\bfseries},
    column{7} = {font=\bfseries},
}

    \abthead{Variant} & \abthead{Iter.} & \abthead{Lat.\\($10^7$)} & \abthead{Energy\\($10^{11}$)} & \abthead{MC} & \abthead{Obj.\\($10^{20}$)} & \abthead{Norm.} \\

    Beacon & 57 & 2.54 & 2.63 & 48.71 & 3.250 & 1.000 \\
    w/o ReAct & 95 & 2.52 & 2.72 & 55.12 & 3.768 & 1.159 \\
    w/o RAG Memory & 96 & 2.38 & 2.50 & 71.06 & 4.229 & 1.301 \\    
    Single Agent & 65 & 2.91 & 3.11 & 53.90 & 4.873 & 1.499 \\
\end{tblr}

\end{table}

We use the 512TOPS task as the benchmark for ablation analysis to evaluate the effectiveness of the hierarchical agent design, RAG Memory, and the ReAct paradigm. Table~\ref{tab:ablation} shows the results.

\subsubsection{Single Agent vs. Multi-Agent}
\label{subsubsec:single_vs_multi_agent}

The single-agent variant achieves an objective of $1.499{\times}$ that of Beacon. It selects a hardware similar to that of Beacon, but increases the DRAM bandwidth to 32 and reduces the GLB capacity of one WS chiplet to 4096. The smaller local buffer limits data residency, while the higher DRAM bandwidth increases hardware cost. Consequently, its latency, energy, and MC are all higher than those of Beacon. This result shows that the single-agent approach can identify parameters that require adjustment, but has difficulty validating the interactions among multiple parameter changes while simultaneously handling global bottlenecks, layer-level behavior, and hardware decisions. Among the three ablations, the single-agent variant causes the largest objective degradation, indicating that the hierarchical multi-agent design contributes the most to search quality.

\subsubsection{Effect of RAG Memory}
\label{subsubsec:rag_memory_ablation}

After removing RAG Memory, the objective becomes $1.301{\times}$ that of Beacon. The resulting design uses eight WS chiplets and retains both the NoP bandwidth and micro-batch size at 64. The larger micro-batch size and higher communication bandwidth reduce latency and energy to $2.38\times10^7$ and $2.50\times10^{11}$, respectively. However, the NoP and related resource configurations increase MC to 71.06, offsetting the gains in the other two metrics.

\subsubsection{Effect of the ReAct Paradigm}
\label{subsubsec:react_ablation}

In this ablation, each agent selects all tool calls at once based on the initial context, rather than dynamically making subsequent decisions based on the results returned by individual tool calls. The best design found by this variant has an objective of $1.159{\times}$ that of Beacon. It still uses eight chiplets, TP8, and a micro-batch size of 32, but changes the dataflow composition to four WS and four OS chiplets and increases the DRAM bandwidth to 32. As a result, energy and MC increase to $2.72\times10^{11}$ and 55.12, respectively, demonstrating that the ReAct paradigm improves the effectiveness of search decisions.

\subsection{Overhead Analysis}
\label{subsec:overhead_analysis}

\begin{table}[t]
    \centering
    \caption{Runtime breakdown of the 512 TOPS Beacon run. Avg./Iter. denotes the average time consumed per search iteration.}
    \label{tab:overhead}
    \scriptsize
    \begin{tblr}{
        width = \columnwidth,
        colspec = {Q[c,m,wd=0.3\columnwidth] X[c,m] X[c,m] X[c,m]},
        cells = {halign=c, valign=m},
        row{1} = {font=\bfseries},
        hlines,
        vlines,
    }
        Component & Total (h) & Avg./Iter. (s) & Share \\
        Model-Level Agent & 2.42 & 87.0 & 5.3\% \\
        Layer-Level Agent & 3.23 & 116.4 & 7.1\% \\
        Solution Agent & 5.96 & 214.4 & 13.1\% \\
        Evaluator & 34.07 & 1214.5 & 74.5\% \\
        RAG Memory & 0.00 & 0.2 & 0.0\% \\
    \end{tblr}
\end{table}

We analyze the overhead of Beacon in terms of runtime and API cost.
Table~\ref{tab:overhead} reports the runtime breakdown of the main 512-TOPS
experiment. The evaluator dominates the overall HW-DSE process,
accounting for 74.5\% of the total runtime. Among the three agents,
the Solution Agent incurs the largest total runtime overhead, although
it does not make the largest number of calls, as shown in Fig.~\ref{fig:llm_backend}(d).
Its average call time reaches 27.7 seconds, higher than those of the
other two agents. This is because the final solution-generation stage
must integrate bottleneck summaries, historical cases, the hardware
search space, and candidate actions, making each call more
computationally demanding. The overhead of RAG Memory itself is
small: one hundred writes and retrievals take approximately
16.4 seconds in total, substantially less than the overhead of the
other components.

API cost is mainly determined by token usage and cache hit rate. The 512TOPS run includes 2270 LLM API calls and about 35.9M tokens in total. The prompt tokens account for 33.0M, and output tokens account for 2.9M. DeepSeek reports 27.7M prompt-cache-hit tokens, corresponding to a cache hit rate of 84.0\%. The estimated API cost of this run is about \$4.94, mainly from cache-miss input tokens and output tokens.

Overall, Beacon trades LLM analysis overhead for higher-quality evaluation decisions. The decision overhead remains on the same order as the simulator evaluation time. This allows Beacon to obtain better designs than traditional methods within tens to hundreds of design evaluations.

\section{Conclusion}

In this work, we propose Beacon, an LLM-based multi-agent HW-DSE framework for heterogeneous multi-chiplet accelerators. Beacon targets the challenge of making effective hardware search decisions under a limited evaluation budget. It turns fine-grained evaluation reports into structured optimization guidance, and uses hierarchical agents, an analysis toolbox, and RAG Memory to localize bottlenecks, diagnose root causes, and generate hardware modifications. Experiments on large-model inference workloads show that Beacon significantly reduces the joint latency, energy, and monetary-cost objective over random search, Bayesian optimization, and reinforcement learning, while discovering effective heterogeneous dataflow designs.

\section{Acknowledgement}


During the preparation of this manuscript, generative AI tools were used for language editing and grammar refinement. DeepSeek-V4-Pro, Gemini-3.1-Pro, GPT-5.5, and Claude-Opus-4.8 were also evaluated as experimental backends in the proposed HW-DSE framework. All technical contributions, results, and conclusions were developed and verified by the authors, who take full responsibility for the manuscript.

\bibliographystyle{IEEEtran}
\bibliography{ref}
\vspace{-15mm}

\begin{IEEEbiography}[{\includegraphics[width=1in,height=1.25in,clip,keepaspectratio]{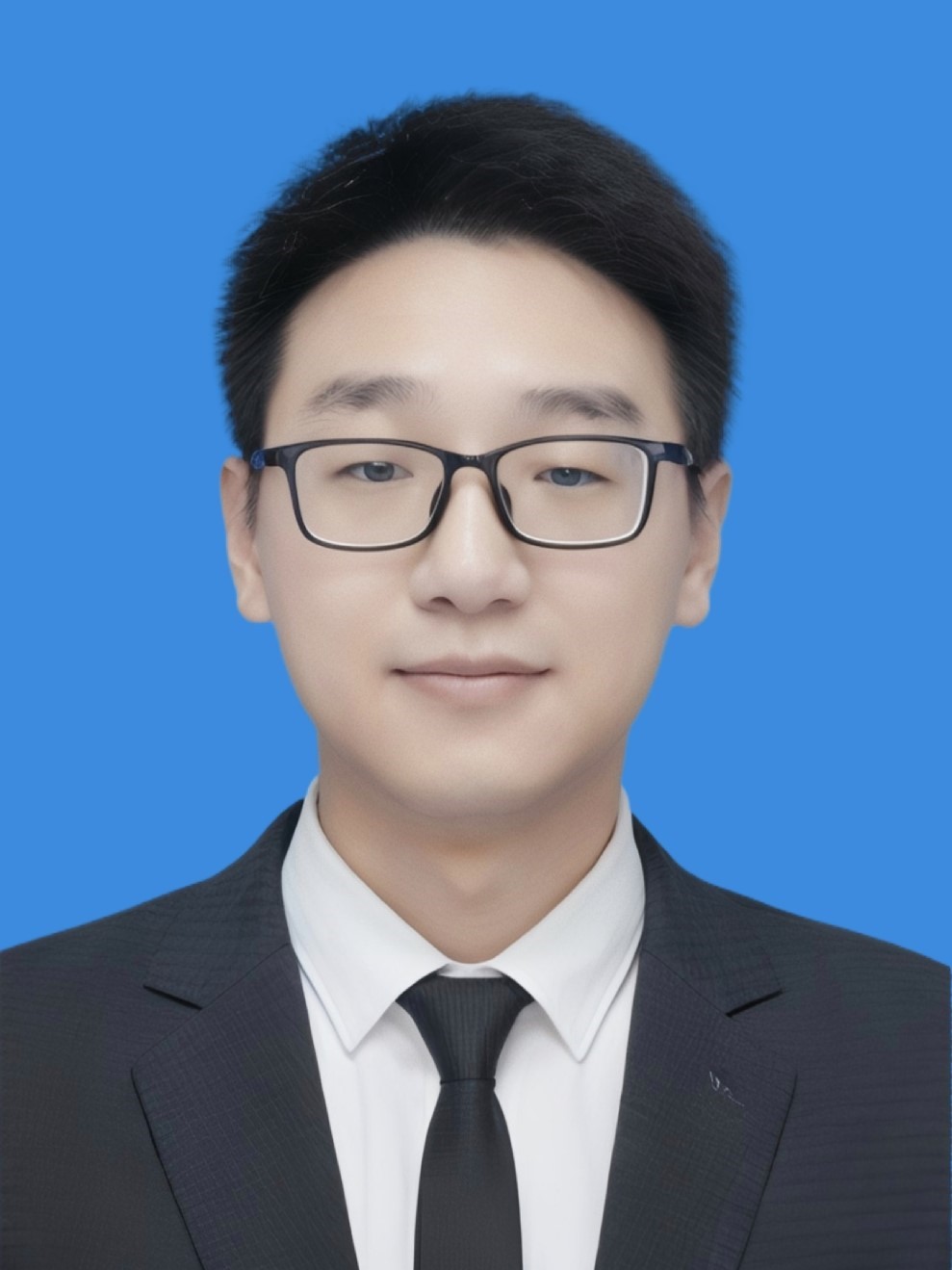}}]{Boyu Li}
received his B.S. degree in computer science and technology from Tongji University, in 2022. He is currently working toward the Eng.D degree in the School of Computer Science and Technology at the University of Science and Technology of China (USTC). His research interests include deep learning accelerator, multi-chiplet architecture, and inference optimization.\end{IEEEbiography}
\vspace{-11mm}

\begin{IEEEbiography}[{\includegraphics[width=1in,height=1.25in,clip,keepaspectratio]{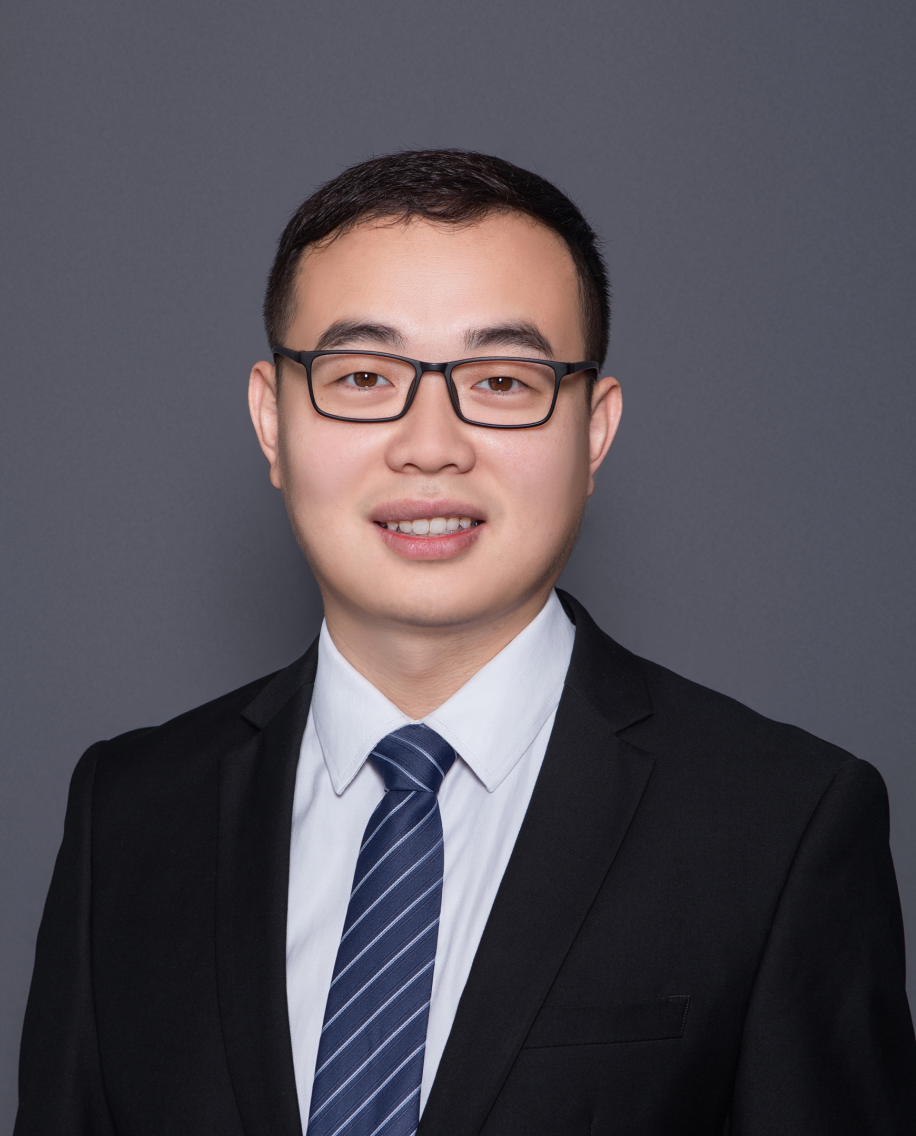}}]{Zongwei Zhu}
received his M.S. and Ph.D. degrees in Computer Science from the University of Science and Technology of China (USTC), in 2011 and 2014, respectively. From 2014 to 2016, he was a research assistant at the China University of Mining and Technology. From 2016 to 2018, he worked as a senior engineer at Huawei Company. Currently, he is an associate professor at the Suzhou Institute for Advanced Research of USTC. His research interests include AI architecture, edge computing, Low-power control and operating system.\end{IEEEbiography}
\vspace{-11mm}

\begin{IEEEbiography}[{\includegraphics[width=1in,height=1.25in,clip,keepaspectratio]{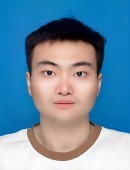}}]{Qianyue Cao} received the B.S. degree in computer science and technology from the Hefei University of Technology, Hefei, China, in 2021. He is currently pursuing Eng.D. in computer science at the School of Computer Science, University of Science and Technology of China, Hefei. His research focuses on edge computing and heterogeneous efficient computing.\end{IEEEbiography}
\vspace{-11mm}

\begin{IEEEbiography}[{\includegraphics[width=1in,height=1.25in,clip,keepaspectratio]{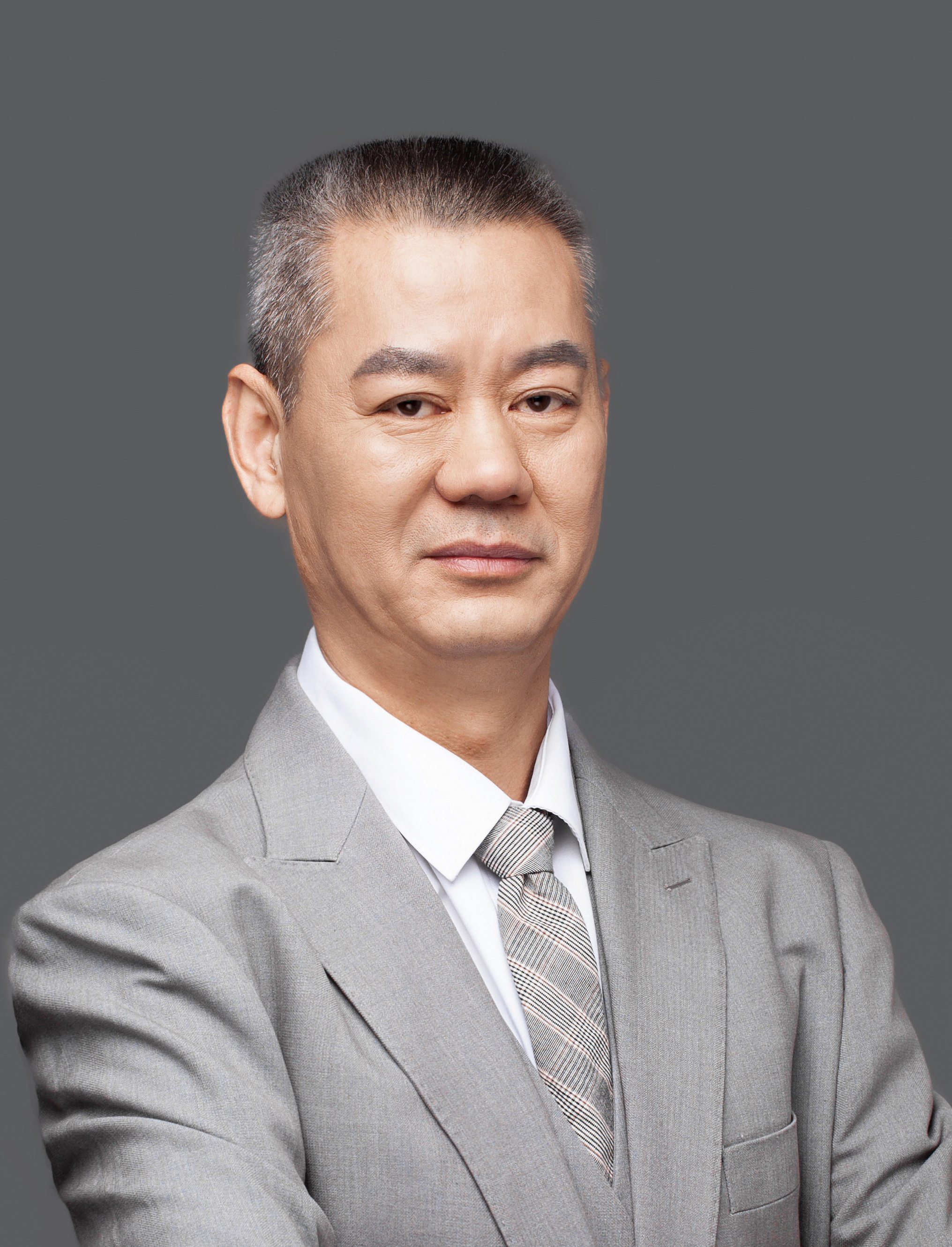}}]{Xi Li} received the Ph.D. degree in computer science from the University of Science and Technology of China, in 2003. He is currently a professor in the School of Computer Science and Technology, and the School of Software Engineering, University of Science and Technology of
China. There he directs the research programs in
High Energy-efficiency Intelligent Computing
Lab, examining various aspects of computer systems, especially real-time embedded systems.\end{IEEEbiography}
\vspace{-11mm}

\begin{IEEEbiography}[{\includegraphics[width=1in,height=1.25in,clip,keepaspectratio]{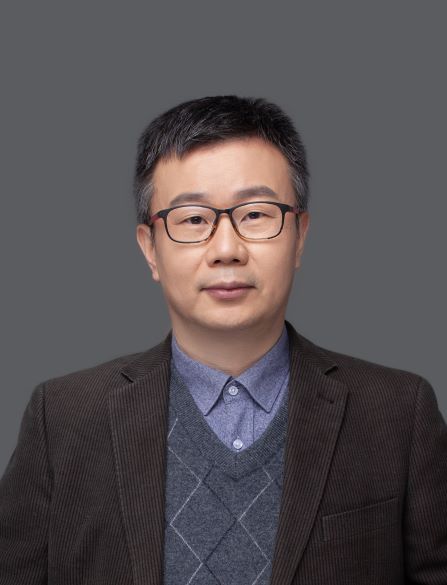}}]{Xuehai Zhou} (Member, IEEE)  received the B.S., M.S., and Ph.D. degrees in computer science from
the University of Science and Technology of China (USTC), Hefei, China, in 1987, 1990, and 1997, respectively. He is a Professor with the School of Computer Science and the School of Software Engineering, USTC. His research interest includes various aspects of multicore and distributed systems. Prof. Zhou serves as a General Secretary of the Steering Committee of Computer College Fundamental Lessons, and the Technical Committee of Open Systems, CCF.
\end{IEEEbiography}

\end{document}